\documentclass[preprint]{aastex}

\slugcomment{To appear in June 10, 2001 issue of ApJ}

\shorttitle{Heckman et al.}
\shortauthors{OVI in NGC1705}

\begin{document}


\title{FUSE Observations of Outflowing OVI in the Dwarf Starburst Galaxy
NGC~1705}


\author{T. M. Heckman\altaffilmark{1,2},
K. R. Sembach\altaffilmark{1}, G. R. Meurer\altaffilmark{1},
and D. K. Strickland\altaffilmark{3}}
\affil{Department of Physics \& Astronomy, Johns Hopkins University,
Baltimore, MD 21218}

\author{C. L. Martin\altaffilmark{1}}
\affil{Department of Astronomy, Caltech, Pasadena, CA 91125}

\and

\author{D. Calzetti\altaffilmark{1} and C. Leitherer\altaffilmark{1}}
\affil{Space Telescope Science Institute, Baltimore, MD 21218}


\altaffiltext{1}{Guest Investigators on the NASA-CNES-CSA Far
Ultraviolet 
Spectroscopic Explorer. FUSE is operated for NASA by the Johns Hopkins 
University under NASA contract NAS5-32985.} 
\altaffiltext{2}{Adjunct Astronomer, Space Telescope Science Institute}
\altaffiltext{3}{Chandra Fellow}


\begin{abstract}
We report $FUSE$ far-UV spectroscopy of
the prototypical dwarf starburst galaxy
NGC~1705. These data allow us for the first time to directly probe
the coronal-phase ($T$ = few $\times$10$^5$ K) gas that may dominate
the radiative cooling of the supernova-heated ISM and thereby determine
the dynamical evolution of the starburst-driven outflows in dwarf
galaxies. We detect a broad
($\sim$100 km s$^{-1}$ FWHM) and blueshifted ($\Delta$$v$ = 77 km s$^{-1}$)
$OVI\lambda$1032 absorption-line arising in the
previously-known galactic outflow.
The mass and kinetic energy in the outflow we detect
is dominated by the warm ($T \sim$ 10$^4$ K)
photoionized gas which is also seen through its optical line-emission.
The kinematics of this warm gas are compatible
with a simple model of the adiabatic expansion of a superbubble
driven by the collective effect of the kinetic energy supplied
by supernovae in the starburst.
However, the observed properties
of the $OVI$ absorption in NGC~1705 are {\it not} consistent with
the simple superbubble model, in which the $OVI$ would arise in a conductive
interface inside the superbubble's outer shell.
The relative outflow speed of the $OVI$ 
is too high and the observed column density ($logN_{OVI}
=$ 14.3)
is much too large. We argue that the superbubble
has begun to blow out of the ISM of NGC~1705. During this 
blow-out phase the superbubble shell accelerates and fragments. The
resulting hydrodynamical interaction as hot outrushing gas flows between
the cool shell fragments will create intermediate-temperature
coronal gas that can produce the observed $OVI$ absorption.
For the observed flow speed of $\sim10^2$ km s$^{-1}$,
the observed $OVI$ column density is just what is expected for
gas that has been heated and which then cools radiatively.
Assuming that the coronal-phase gas is in rough pressure balance with
the warm photoionized gas, we estimate a cooling rate of-order
$\sim$0.1 $M_{\odot}$ per year and $\sim$10$^{39}$ erg s$^{-1}$ in the
coronal gas. The latter represents $<$10\% of the supernova
heating rate. Independent of the assumed pressure,
the lack of observed redshifted $OVI$ emission from
the backside of the outflow leads to upper limits on the cooling
rate of $\leq$ 20\% of the supernova heating rate.
Since the X-ray luminosity of NGC~1705 is negligible,
we conclude that radiative-losses are insignificant in the outflow.
The outflow should therefore be able to fully blow-out of the ISM
of NGC~1705 and vent its metals and kinetic energy. This process
has potentially important implications for the evolution of dwarf
galaxies and the IGM.
\end{abstract}




\section{Introduction}

Local starburst galaxies are excellent local laboratories
to study the physics of galaxy building (cf. Heckman 1998).
It is now clear that the deposition of mechanical and thermal energy
by multiple supernovae in starbursts leads to a global outflow of
metal-enriched gas. These outflows are called ``superbubbles'' during
their early dynamical 
evolution, and ``superwinds'' after they blow out of the galaxy's ISM
(e.g., Heckman 2000 and references therein). Such flows are expected
to play an especially important role in the evolution of dwarf galaxies,
whose relatively shallow potential wells make them susceptible
to wind-driven loss of gas and newly-created metals (e.g., Dekel \& Silk
1986; Martin 1999). Since similar outflows appear to be common in
high-redshift galaxies as well (Pettini et al. 1998, 2000;
Tenorio-Tagle et al. 1999)
they are perhaps the most plausible mechanism by which the mass-metallicity
relation in galactic spheroids was established (e.g., Lynden-Bell
1992) and the intergalactic medium was heated and chemically-enriched
(e.g., Ponman, Cannon, \& Navarro 1999; Gibson, Loewenstein, \& Mushotzky 1997).

One of the major uncertainties concerning starburst-driven outflows
is the importance of radiative cooling: what fraction of the kinetic
energy supplied by supernovae is carried out in the flow rather
than being radiated away? 
While the available X-ray data and models establish that
radiative losses from hot gas ($T>10^6$K) are not severe 
(see Strickland \& Stevens 2000),
up until now there has been no direct observational probe of the
coronal-phase gas ($T = 10^5$ to $10^6$ K)
that could dominate the radiative cooling.

The recent launch of the $Far~Ultraviolet~Spectroscopic~Explorer$
($FUSE$ - Moos et al. 2000) provides access to the best probe of
rapidly-cooling coronal gas in starburst outflows:
the OVI$\lambda$$\lambda$1032,1038 doublet.
Accordingly, we have obtained $FUSE$ spectra of a small sample of
the nearest and brightest starbursts. In this paper, we present
the first $FUSE$ detection of $OVI$ in the galaxy
NGC~1705.
 
NGC~1705 is an ideal test-case for the study of coronal-phase gas.
This nearby (D = 6.2 Mpc) dwarf starburst
galaxy has the second highest vacuum-ultraviolet flux
of any starburst galaxy in the extensive compilation of IUE spectra in Kinney
et al. (1993). The detailed investigation by Meurer et al. (1992) 
established NGC~1705 as a prototypical example of a dwarf starburst
undergoing mass-loss. They were able to delineate a kpc-scale
fragmented ellipsoidal shell of emission-line gas that was expanding
at roughly 50 km s$^{-1}$ along our line-of-sight. They also showed that the
population of supernovae in the young super star cluster (NGC~1705-1)
was energetically-sufficient to drive this flow. The expulsive nature
of the flow was later confirmed by $HST$ observations that showed that the
ultraviolet interstellar absorption-lines towards NGC~1705-1
were blueshifted by 70 to 80 km s$^{-1}$ relative to the galaxy systemic
velocity (Heckman \& Leitherer 1997; Sahu \& Blades 1997; 
Sahu 1998). Hensler et al. (1998)
reported the detection of soft X-ray emission, presumably from hot gas
inside the expanding emission-line nebula.

\section{Observations \& Data Reduction}

Two $FUSE$ observations of NGC\,1705 ($\alpha_{2000} = 04^h 54^m 13.48^s, 
\delta_{2000} = -53^o 21^m 39.4^s$; $l_{II} = 261.0788^o, b_{II} = -38.7428^o$)
were obtained on 4-5 February 2000.
The central super star cluster NGC~1705-1 was centered in the large 
(LWRS, $30\arcsec\times30\arcsec$) aperture of the LiF1 (guiding) channel 
for each observation by the standard guide-star acquisition procedure.  
The two observations resulted in 13 exposures totaling 21.3 $ksec$ of 
on-target exposure time. 
Approximately 95\% of the observing time occurred during orbital night, which
greatly reduced the amount of terrestrial \ion{O}{1} and \ion{N}{1} airglow 
entering the FUSE apertures.
Flux was recorded through the LWRS
apertures in both long 
wavelength (LiF, $\sim1000-1187$\,\AA) channels and both short wavelength
(SiC, $\sim900-1100$\,\AA) channels. The astigmatic heights of the LiF1
spectra near 1030\,\AA\ were roughly 1/3 of the aperture width, consistent 
with the compact photometric structure of NGC~1705.
\footnote{
$HST$ images at 2200\AA\ show that roughly
40\% of the light that would be admitted into the $FUSE$ LWRS aperture
would come from the single brightest star cluster NGC~1705-1, and 70\%
of the light would come from the region within a radius of 2.8 arcsec
of NGC~1705-1 (Meurer et al 1995). A comparison of
the HST $GHRS$ (1.7 arcsec aperture) and $IUE$ (10 by 20 arcsec aperture)
spectra
implies that these fractions should not be strongly wavelength dependent.
Thus, while the $FUSE$ data sample the entire
inner region of NGC~1705 (projected size of roughly 900 by 900 pc),
the sightlines within about 80 pc of NGC~1705-1 are very strongly
weighted.} The data are preserved in
the $FUSE$ archive with observation identifications A0460102 and A0460103.

The raw time-tagged photon event lists for each exposure were processed
with the standard $FUSE$ calibration software ($CALFUSE~v1.7.5$) available at 
the Johns Hopkins University as of August 2000.  The lists were screened 
for valid data, and corrections for geometric distortions, spectral motions, 
and Doppler shifts were applied (see Sahnow et al. 2000).
The data were not affected by the detector event bursts that plagued many 
of the early FUSE spectra.
The 13 individual
calibrated extracted spectra for each channel were cross-correlated, shifted
to remove residual velocity offsets due to image motion in the apertures,
and combined to a produce composite spectrum for each channel. 
These composite spectra in
the 1000--1070\,\AA\ region were then compared, and any remaining velocity 
offsets were removed by referencing to the LiF1 channel data.  The wavelengths
were put into the 
Local Standard of Rest (LSR) reference frame by requiring that the 
Galactic ISM lines fall at $\sim+20$ km~s$^{-1}$,
the approximate velocity of the Milky Way lines as observed at longer
wavelengths with the Space Telescope Imaging Spectrograph (STIS) on the 
$Hubble Space Telescope$.  These new STIS data will be presented in future 
papers (Sembach et al. 2001; Heckman et al 2001).
The $FUSE$ data have a wavelength uncertainty of 
$\sim6$ km~s$^{-1}$ ($1\sigma$).
The spectrum of NGC~1705 is shown in Figure 1.

In this article, we do not combine the data from all four channels because
the instrumental resolution and sensitivity changes as a function of 
wavelength.  Therefore, when measuring the strengths of absorption features, 
we compare the individual measurements (W$_\lambda$) for the two channels 
having the highest sensitivity at the wavelengths of the absorption features 
of interest (usually LiF1 and LiF2 for $\lambda > 1000$ \AA, or SiC1 and 
SiC2 for $\lambda < 1000$ \AA).  The integrated equivalent widths derived 
from separate channels generally agree very well.  Table~1 contains these 
measurements for selected Milky Way and NGC\,1705 absorption features 
observed along the sight line.  The data have a velocity resolution of 
$\sim30$ km~s$^{-1}$ and S/N ratios of 16, 13, 10, and 9 per resolution 
element at 1032 \AA\ in the LiF1, LiF2, SiC1, and SiC2 channels, respectively.

\begin{figure}
\plotone{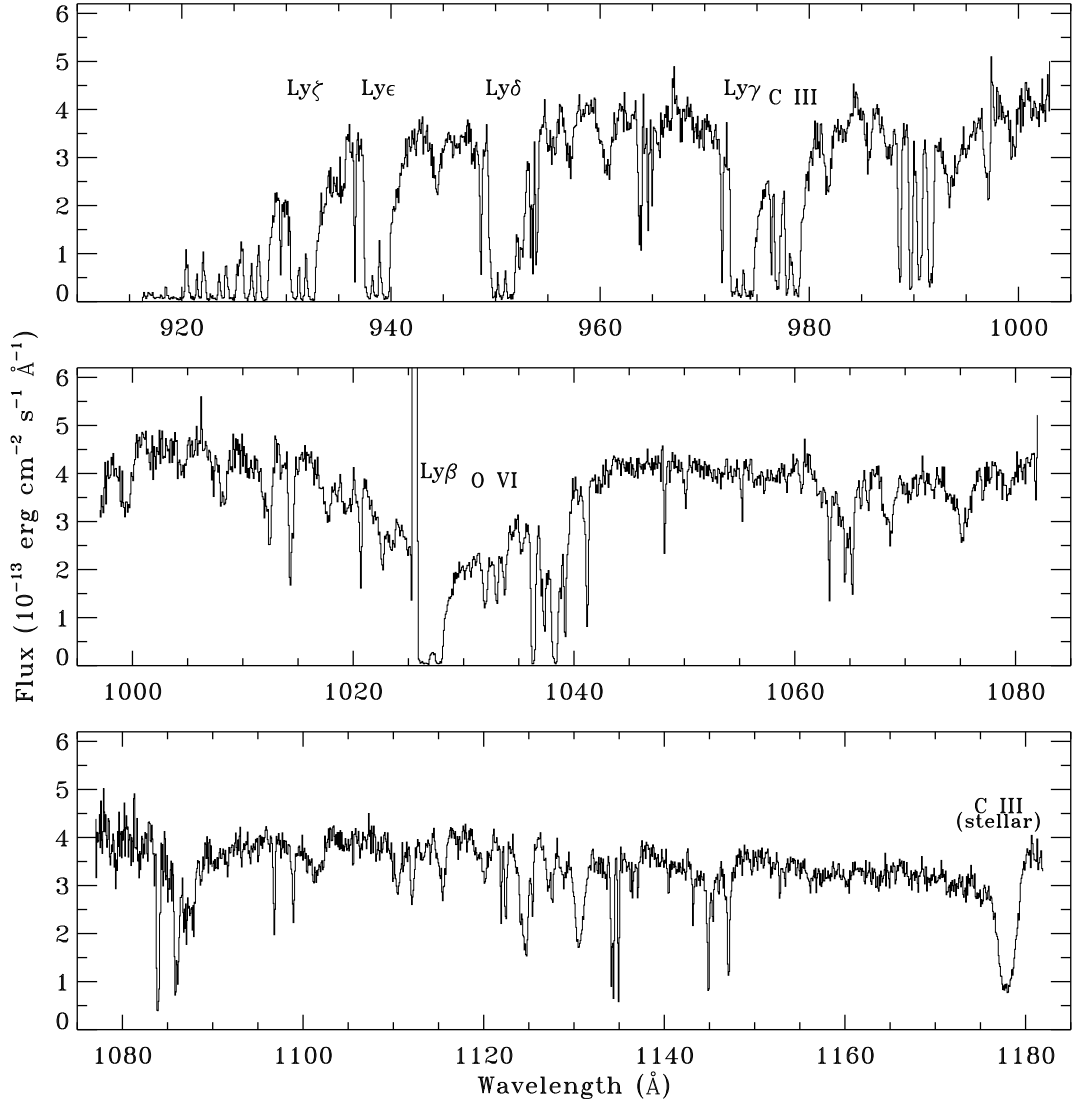}
\vskip -5cm
\caption{Overview of our $FUSE$ spectrum on NGC~1705.
These data have been binned
into 0.04\,\AA\ samples for clarity.  Data from single segments are
shown (SiC2A: 910--1005\,\AA, LiF1A: 995--1085\,\AA,
SiC2B: 1075--1090\,\AA, LiF2A: 1090--1182\,\AA).
The convergence of the
Lyman series of hydrogen is evident, as are numerous strong interstellar
lines arising in both the Milky Way and NGC~1705. The upward spike
near 1026\,\AA\ is due to $Ly\beta$ airglow.}
\end{figure}

\begin{figure}
\plotone{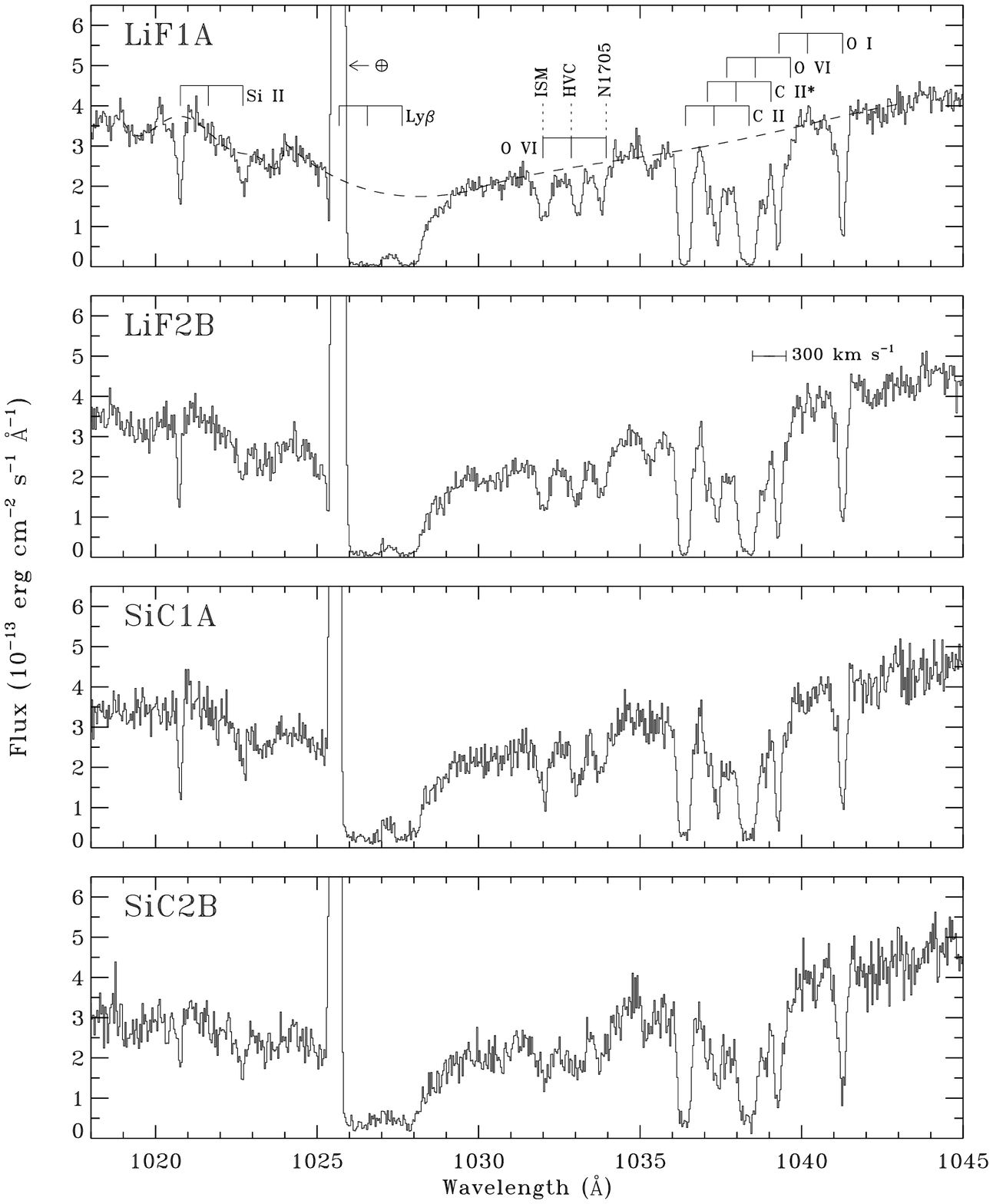}
\vskip -35mm
\caption{$FUSE$ data for the four detector segments covering the
1018--1045\,\AA\
spectral region. These data have been binned into 0.04\,\AA\ samples.
 Prominent
absorption lines of atomic species in the ISM of the Milky Way, HVC~487,
and NGC~1705 are marked above the LiF1A spectrum.  The HVC is seen only
in the lines of $Ly\beta$, $OVI$, and $CII$ in the spectral region shown.
$HI$ $Ly\beta$ airglow
present near zero velocity can be used to judge the spectral resolution
for a source filling the LWRS apertures (FWHM $\sim120$ km~s$^{-1}$).
The data have a (point source) spectral resolution of $\sim30$ km~s$^{-1}$
and S/N $\sim$ 20--25 per resolution element in the wavelength region
shown.  The dashed line over-plotted on the top spectrum indicates the
continuum adopted for our analyses of the absorption lines.}
\end{figure}

\section{Results}

We confirm the principal conclusion of Sahu \& Blades (1997) and Sahu (1998):
there are three separate systems of interstellar absorption-lines
along the sight-line to NGC~1705 (Table 1; Figure 2). These arise in the Milky
Way, the
periphery of the high velocity cloud HVC~487, and the interstellar medium
of NGC~1705. In the present paper, we will focus our analysis on
NGC~1705, but briefly summarize our results on HVC~487.

\subsection{HVC~487}

As pointed out by Sahu \& Blades (1997), HVC~487
is located 2$^{\circ}$ away from the NGC~1705 sight-line. Sahu (1998)
proposes that HVC~487 is associated with the Magellanic Stream, so
the implied impact parameter would be roughly 2 kpc.
The properties of the absorption associated with HVC~487 will be discussed
in detail in a future paper which will combine the $FUSE$ data with echelle
spectra taken with STIS on $HST$ (Sembach et al. 2001). Here, we note only that
HVC~487 is strongly detected in absorption in the
$OVI\lambda$1032, $CIII\lambda$977,
$CII\lambda$1036, and Lyman-series lines. Assuming
that the $OVI\lambda$1032 line is optically-thin, the implied $OVI$
column is 2.0$\pm$0.3 $\times$ 10$^{14}$ cm$^{-2}$. This is typical of other
HVC sight-lines studied with $FUSE$ (Sembach et al. 2000). 
The Lyman series, CIII$\lambda$977, and CII$\lambda$1036 absorption-lines
from HVC~487 are at $v_{LSR}$ = 270$\pm$15 km s$^{-1}$. These are close to the 
$HI\lambda$21cm velocity of HVC~487 of 232$\pm$14 km s$^{-1}$
(Sahu 1998). In contrast,
the $OVI\lambda$1032 line is centered at $v_{LSR}$
= 326$\pm$10 km s$^{-1}$, an offset of 94$\pm$17 km s$^{-1}$ from the 
$HI\lambda$21cm velocity.
The $OVI$ line is very broad, with an observed
FWHM = 100$\pm$15 km s$^{-1}$. This breadth is comparable to
the sound speed in coronal gas where $OVI$ would be abundant,
and is similar to those of other $OVI$ HVCs observed with $FUSE$
(Sembach et al. 2000).

\subsection{NGC~1705}

We expect our $FUSE$ data to probe four phases of interstellar gas
in NGC~1705: molecular gas, neutral atomic gas, warm gas photoionized
by hot stars, and collisionally-heated coronal gas.
The neutral atomic gas is traced
by species with ionization potentials of creation $\chi<$ 1 Ryd, the warm gas
by ions with $\chi$ = 1 to 4 Ryd (up to the HeII edge), and
the coronal gas by ions with $\chi>$4 Ryd.

No molecular
hydrogen was detected in any of the first three $J$ levels (0-2),
and the limit on the total $H_2$ column density is $logN_{H2} <$ 14.6
(corresponding to a molecular hydrogen fraction
of $f_{H2} = 2N_{H2}/(N_{HI}+2N_{H2}) < 6\times10^{-6}$
 - based on the observed HI column given below). Similar results
have been reported for the metal-poor dwarf starburst IZw18
(Vidal-Madjar et al. 2000). The intrinsic reddening in NGC~1705 is very small:
Heckman et al. (1998) give $E(B-V) <<$ 0.1.  
In the particular case of NGC~1705,
the low values for $N_{H2}$ and $f_{H2}$ are quite consistent with
Galactic sightlines with similarly-low $E(B-V)$ and
$N_{HI+H2}$ (Savage et al. 1977).

Lines tracing the other
three phases are well-detected (Table 1; Figures 1 and 3).
In particular, the $OVI\lambda$1032
line is independently detected at the 10$\sigma$ level 
in both the LiF1 and LiF2 detectors (Figure 2). The weaker
member of the doublet ($OVI\lambda$1038) is blended with the
much stronger Galactic $OI\lambda$1039 line and is not convincingly
detected.

\begin{figure}
\includegraphics[height=18cm]{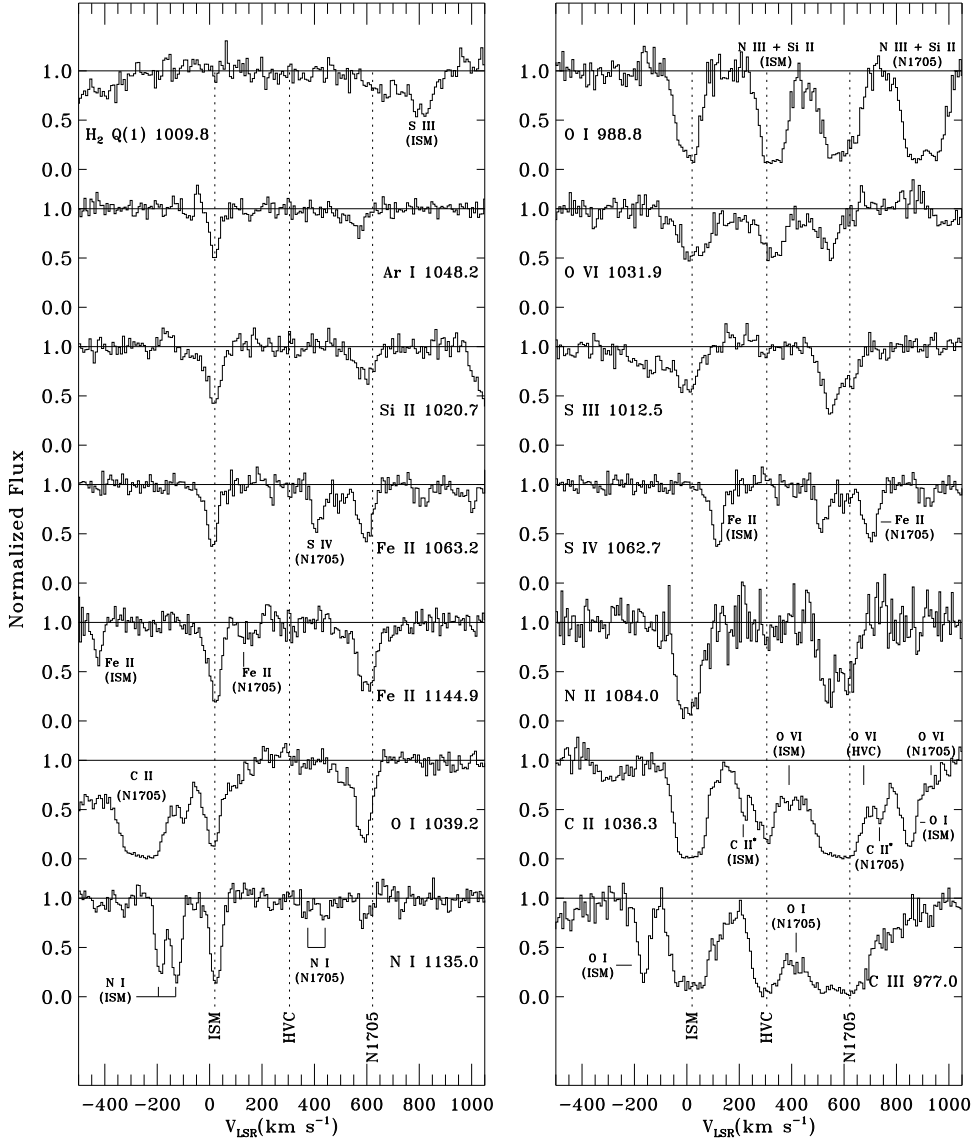}
\caption{Continuum-normalized absorption line profiles versus LSR
velocity for
selected absorption lines in the $FUSE$ bandpass.
The profiles shown are from the LiF1 channel, except for CIII
$\lambda977.0$, OI$\lambda988.8$, and NII$\lambda1084.0$,
which are from the SiC2 channel.  Data of roughly comparable quality exists
for the LiF2 and SiC1 channels (see Table~1).
The identifications of
the lines are listed underneath each spectrum.  Additional lines arising
in the ISM or within NGC~1705 within the velocity range shown are also
indicated (without wavelengths).  In some cases, lines from various
species blend and cannot be resolved.  The vertical dashed lines indicate
the LSR velocities of the ISM
(20 km~s$^{-1}$), HVC~487 ($\sim305$ km~s$^{-1}$
compared to the $HI\lambda$21cm emission velocity of 232 km~s$^{-1}$),
and the systemic velocity of NGC~1705 (622 km~s$^{-1}$).  Note that the
NGC~1705 absorption lines are substantially blue-shifted relative to the
systemic velocity of the galaxy.}
\end{figure}

\subsubsection{Kinematics}
We begin by considering the kinematic properties of the gas.
In what follows, we
will adopt an LSR systemic velocity for NGC~1705 of 
$v_{sys}$ = 622 km s$^{-1}$, which is 
derived from the HI$\lambda$21cm rotation curve (Meurer et al. 1998).

The mean radial velocity of the neutral gas lines listed in Table 1 is 
590$\pm$11 km~s$^{-1}$ as determined by fitting Gaussian components to the 
observed absorption features. This implies a blue-shift of 32 km~s$^{-1}$
relative to $v_{sys}$, which is significantly larger than the uncertainty 
11 km~s$^{-1}$ derived from the standard deviation of the line centroids.
In contrast, the coronal gas traced by $OVI$ is significantly
more blueshifted: $v_{OVI}$ = 545$\pm$10 km s$^{-1}$, or $v - v_{sys}$ =
-77 km s$^{-1}$. The warm photoionized gas appears to have
intermediate velocities with the mean line centroid at 569$\pm$10
km s$^{-1}$ ($v - v_{sys}$ = -53 km s$^{-1}$)\footnote{We exclude
the highly saturated and very broad $CIII\lambda$977 line, which
is discussed separately}.

Further evidence for differing dynamics between the neutral and coronal
gas comes from the line widths (see Figure 3). The weak (unsaturated) lines from
the neutral phase have observed lines widths (FWHM) of
roughly 70 km s$^{-1}$, while the $OVI\lambda$1032 line has a broader
structure (FWHM $\sim$ 100$\pm$10 km s$^{-1}$). The lines from the photoionized
gas are roughly the same width as the $OVI$ line.

The strongly saturated $CIII\lambda$977 and $CII\lambda$1036 lines
allow us to probe the low-column density gas at extreme radial
velocities. Relative
to $v_{sys}$, the centroid of the $CIII$ line is blueshifted
by 95$\pm$20 km s$^{-1}$. The blueward edge of the line is blended
with the HVC feature, but absorption extends to at least
-260 km s$^{-1}$ with respect to $v_{sys}$.
Relative to $v_{sys}$,
the $CII\lambda$1036 line centroid
is blueshifted by 40$\pm$20 km s$^{-1}$ km s$^{-1}$ and 
absorption is present from at least -180 km s$^{-1}$ (where the line is blended
with the HVC feature) to roughly +70 km s$^{-1}$. 
The HI$\lambda$21cm maps of NGC~1705 published by Meurer et al (1998) show
emission
from -94 to +86 km s$^{-1}$ relative
to $v_{sys}$. The velocity range of the {\it redshifted}
gas seen in $CII\lambda$1036 is thus consistent with the kinematics
of the normal neutral (turbulent?) ISM, but an additional
component of highly blueshifted (outflowing) neutral gas is also present.

\subsubsection{Column Densities and Abundances}

We have determined column densities by converting the observed
absorption profiles into optical depth profiles and integrating
over velocity (see Savage \& Sembach 1991). We measured only the weaker
(and thus, more-optically-thin) metal lines
with equivalent widths $W <$ 250 m\AA. The results are given in Table 2.

For the lines in the neutral
phase, we estimate the following column densities (log$N$):
$NI$ (14.0), $OI$ (15.6), $SiII$ (14.6), $ArI$ (13.5), and $FeII$ (14.5).
Heckman \& Leitherer (1997) estimated that 
$N_{HI}$ = 1.5$\times$10$^{20}$ cm$^{-2}$
column towards NGC~1705-1 based on fitting the red side of the
damped $Ly\alpha$ profile in the HST $GHRS$ data. We have fit the Ly$\beta$
and higher-order
Lyman series lines in our $FUSE$ spectrum, and find
$N_{HI}$ = 2$\pm$1 $\times$10$^{20}$ cm$^{-2}$
($b = 35\pm$5 km s$^{-1}$).
We adopt the mean of these two estimates:
log$N_{HI}$ = 20.2$\pm$0.2.
Since these species above should be the dominant state
of their respective element in the neutral gas, we can
estimate the following abundances
(Anders \& Grevesse 1989):
[N/H]=-2.2, [O/H]=-1.5, [Si/H]=-1.1, [Ar/H]=-1.3, and [Fe/H]=-1.2.
The absolute abundances are uncertain by 0.3 dex.
With the
exception of nitrogen, the abundances are marginally consistent with 
the metallicity of [O/H]=-0.9 derived for
the nebular
emission-line gas (Heckman et al. 1998). This is
in agreement with expectations that we see ambient interstellar gas,
rather than the much hotter gas that is polluted by recent supernova
ejecta. 

We confirm Sahu`s (1998) finding of a low relative $N$ abundance, based
on the $NI$ column density. As Sahu points out, the $NII$ ion can also make a
significant contribution to the total $N$ column in the $HI$ phase
(see Sofia \& Jenkins 1998).
We estimate log$N_{NII}$ = 14.7. Unless most of this is associated
with the $HI$ phase, $N$ remains selectively underabundant.
In fact, most of the observed $NII$ must be associated with
the warm photoionized gas that dominates the total gas 
column along our line-of-sight (see below). This is implied by
the kinematics of the $NII\lambda$1084 line, which has a significantly larger
blueshift and FWHM than the lines arising in the neutral phase, but agrees
with the other lines arisiing in the warm photoionized gas (Table 1;
and see above).
A low nitrogen abundance
is typical of low-metallicity gas in dwarf galaxies,
and is a consequence of a partly secondary nucleosynthetic origin for $N$
(e.g. van Zee, Haynes, \& Salzer 1997). 
We do not confirm the low relative $Fe$ abundance noted by Sahu \& Blades
(1997). Finally, we see no evidence for the systematic gas-phase depletion
of refractory elements (e.g. compare $Fe$ to $Ar$). The neutral gas
in the outflow thus appears largely dust-free (e.g., Savage \& Sembach
1996), consistent with the
lack of detectable reddening in the UV spectrum of NGC~1705
(e.g., Meurer, Heckman, \& Calzetti 1999).

The total column density in the warm photoionized gas is more uncertain.
The weak $SIII\lambda$1012 and $SIV\lambda$1063 lines
imply column densities log$N_{SIII}$ = 15.0 and log$N_{SIV}$ = 14.4.
The $SII$ ion can also be adundant in photoionized gas.
While the HST $GHRS$ spectrum covers a somewhat different sight-line
(the 1.7$\times$1.7 arcsec Large Science Aperture was centered
on NGC~1705-1), the $SII$ column
density log$N_{SII}$ = 15.0 from Sahu \& Blades (1997) is a useful
additional constraint.
The total implied column of ionized $S$
is log$N_{S}$ = 15.3. For an assumed value [S/H]=-0.9 (based
on nebular emission-line metallicity), the total $HII$ column
in the photoionized gas is log$N_{HII}$ = 20.9, or almost
an order-of-magnitude greater than the $HI$ column.

The observed $OVI\lambda$1032 line yields an $OVI$ column density
log$N_{OVI}$ = 14.3. Deriving the total column density of
the coronal phase gas requires uncertain assumptions. For gas
in collisional ionization equilibrium, $OVI$ reaches its peak
relative abundance ($\sim$20\% of the total $O$ abundance) at
$T\sim$3$\times$10$^5$ K (Sutherland \& Dopita 1993).
If we assume that [O/H]=-0.9 in the coronal
phase gas (similar to the value in the optical emission-line gas),
the implied minimum total $HII$ column in the coronal gas
is then log$N_{cor}$ = 19.0. While uncertain, we note that this
is a substantially smaller column than seen in the cooler gas.

\subsection{Limits on OVI Emission}

In addition to the blueshifted $OVI$ absorption-line produced by
coronal gas on the frontside of the outflow, redshifted $OVI$ {\it emission}
from the backside of the outflow must also be present.
We do not detect such emission, but can set an interesting upper limit
to its intensity.
Combining our LiF1 and LiF2 spectra, we obtain a 3$\sigma$ upper limit 
on the flux of the redshifted
$OVI\lambda$1032 emission-line of $\leq 7.5\times$10$^{-15}$
erg cm$^{-2}$ s$^{-1}$. This limit assumes that the line would have a breadth
similar to the blueshifted absorption-line ($\sim$ 100 km s$^{-1}$).

We first correct this flux for forground dust extinction. The
Galactic $HI$ column towards NGC~1705 is 1.3$\times$10$^{20}$
cm$^{-2}$, based on both the damped $Ly\alpha$ profile (Sahu 1998)
and radio $\lambda$21cm observations (reported in Hensler et al. 1998).
Adopting the standard extinction curve of Mathis (1990), the implied
extinction at 1032\AA\ is 0.3 magnitudes. The data discussed in
section 3.2.2 imply that there is a negligible amount of dust
extinction intrinsic to NGC~1705. 

In the discussion below, we
will be interested in the total $OVI$ luminosity of NGC~1705. To derive
a global upper limit, we make a simple aperture correction to our observed
limit. The portion of the backside of the NGC~1705 outflow included
in our $FUSE$ aperture represents about 15\% of the total surface
area of the outflow (front and back sides), as mapped by Meurer et al. (1992).
Thus, the global extinction-corrected $OVI\lambda$1032 flux from
NGC~1705 is $\leq$ 6.5$\times$10$^{-14}$
erg cm$^{-2}$ s$^{-1}$, and the corresponding luminosity
is $\leq$ 3$\times$10$^{38}$ erg s$^{-1}$.

\section{Discussion}

\subsection{A Simple Model}

The simplest physical model that can be compared to our data
is an adiabatic expanding superbubble
whose expansion is driven by the energy supplied by multiple
supernovae (Weaver et al. 1977; Koo \& McKee 1992). This
is a natural model to use, since it appears to be a good
first-order description of the properties of the $H\alpha$ emission
line nebula in NGC~1705 (Meurer et al. 1992; Marlowe et al. 1995).

In this simple model, there are six concentric zones. From inside-out
these are: $1)$ An innermost region inside which energy is injected 
by supernovae (the starburst). $2)$ A region of supersonic flow
fed by the hot gas created in zone 1. $3)$ A region of hot gas (material
from zone $2$ that has passed through an internal shock). $4)$ A
conductive interface of intermediate-temperature gas created as the
hot gas in zone $3$ heats the
relatively cool, dense gas in zone $5$. $5)$ A thin dense shell
of ambient gas that has been swept-up (shocked and then radiatively cooled)
as the ``piston'' of hot gas in zone $3$ expands into the ISM. Depending
on the available flux of ionizing radiation from the starburst, this
shell may be partially or fully photoionized. $6)$ The undisturbed
ambient ISM.

The dynamical evolution of a superbubble in the type of plane-parallel
ISM appropriate to a starburst has been extensively discussed
(see MacLow \& Ferrara 1999 and references therein). Once the radius
of the superbubble is several times the vertical scale-height
of the ISM, its expansion will accelerate and Rayleigh-Taylor
instabilities will cause the outer shell (zone $5$) to fragment.
This will allow the hot gas from the interior 
to escape from the ruptured superbubble and flow out into the
galactic halo. This ``blow-out'' or ``
break-out'' stage marks the transition
from superbubble to superwind.

\subsection{Some Simple Inferences}

Before considering the superbubble model in more detail, it
is worthwhile to make some rough estimates of the mass and
kinetic energy in the gas and compare these to expectations.
To convert the observed column densities and outflow velocities
into masses and kinetic energy, we will assume the following
idealized model for the NGC~1705 outflow. Meurer
et al. (1992) show that the morphology and kinematics of the
emission-line nebula can be described as the expansion of a
hollow prolate ellipsoid with a semi-major axis of 1.5 kpc
and semi-minor axes of 0.5 kpc. Multiplying the surface area
of this ellipsoid by our estimated column densities (see above)
implies total gas masses of 5$\times$10$^7$, 1$\times$10$^7$,
and 6$\times$10$^5$ $M_{\odot}$ respectively in the warm
photoionized gas, the neutral gas, and the coronal gas. The 
outflow speeds then yield corresponding kinetic energies of 
1.6$\times$10$^{54}$, 1$\times$10$^{53}$, and 3$\times$10$^{52}$ ergs.

The supernovae in the NGC~1705 starburst will 
supply kinetic energy at a mean rate of roughly
2$\times$10$^{40}$ erg s$^{-1}$ (Marlowe, Meurer, \& Heckman 1999).
The dynamical
age of the expanding emission-line nebula is 10 to 15 Myr (Marlowe et al.
1995; Meurer et al. 1992), so the total amount of kinetic energy supplied
by the starburst during this time is $\sim1\times$10$^{55}$ ergs.
In the standard adiabatic superbubble model, the kinetic energy
of the swept-up shell (zone $5$) will be about 19\% of the total
injected energy (e.g. MacLow \& McCray 1988).
Given the nature of our estimates, the rough agreement
between the observed and available/predicted kinetic energies is gratifying.

The amount of mass returned directly by supernovae and stellar winds
during the last 10-15 Myr will only be $\sim$10$^5 M_{\odot}$
(Leitherer \& Heckman 1995). This is tiny compared to
shell's mass (see above), consistent with the basic superbubble
model in which the shell is swept-up ambient ISM. In this case,
the total gas column density seen by $FUSE$ 
(10$^{21}$ cm$^{-2}$) can be no larger than the total
column density of the ISM prior to the expansion of the superbubble.
The HI$\lambda$21cm maps in Meurer et al.
(1998) show that the mean column density in the inner part
of NGC~1705 is about 2$\times$10$^{21}$ cm$^{-2}$, consistent
with this requirement.

The similarity between the total gas column of outflowing gas measured by $FUSE$
and the typical HI column in NGC~1705 implies
that the dimensions of the superbubble must be at least comparable to
the characteristic thickness of the ISM. This suggests that the superbubble
is early in the blow-out stage of it dynamical evolution. Such an
inference is also consistent with the patchy, filamentary morphology of the
superbubble's H$\alpha$ emission
(Meurer et al. 1992; Marlowe et al. 1995), which suggests
that the superbubble shell has begun to accelerate and fragment (see above).

The absorbing $HI$ column in NGC~1705 (log$N_{HI}$ = 20.2$\pm$0.2)
is roughly an order-of-magnitude smaller than the column observed in emission
at $\lambda$21cm in the same region (with the $\sim$30 arcsec $ACTA$ beam
- Meurer et al. 1998). This implies that roughly 90\% of the $HI$ seen
in the radio map lies {\it behind} the dominant source of far-UV light
(the region around the super star cluster NGC~1705-1). That is, we
infer that the superbubble is not symmetrically located within the
HI disk of NGC~1705, but must be on the near side. Thus, the blow-out
is apparently one-sided at present (cf. MacLow, McCray, \& Norman 1989).

\subsection{The Warm Photoionized Gas}

The column 
density of gas along our line-of-sight is dominated by
the warm ionized gas.
This implies that the shell of
swept-up material (zone $5$ in the superbubble) has been mostly photoionized
by Lyman
continuum radiation from the starburst. It is natural to identify
the gas seen in absorption in the $FUSE$ data with
the emission-line
nebula studied by Marlowe et al. (1995) and Meurer et al. (1992).

The kinematics are consistent with this identification.
Recall that the $FUSE$ sight-line is heavily weighted towards
material within about 3 arcsec of the bright super star cluster
NGC~1705-1. The echelle spectra of this region
analyzed by Marlowe et al. (1995) and Meurer et al. (1992) show
double-peaked H$\alpha$ emission-line profiles, with the
blueshifted (redshifted) peak corresponding to emission from
the front (back) side of the expanding superbubble. The measured
LSR radial velocity of the blueshifted component
is 560$\pm$20 km s$^{-1}$ along the different position angles
covered by their combined data sets. Within the uncertainties,
this agrees with the mean velocity for the corresponding
$FUSE$ absorption-lines (569$\pm$10 km s$^{-1}$).

Since we now have a mass for the warm photoionized gas (\S~4.2), its
H$\alpha$ luminosity (Marlowe et al. 1995) can be used
to {\it directly} determine that the mean electron density in this material
is $n_e \sim$ 1 cm$^{-3}$. We emphasize that this makes no assumptions
about the volume filling-factor of the gas. Taking
$T\sim$10$^4$ K (typical of photoionized gas),
the corresponding thermal pressure
is P/k $\sim$ 2$\times$10$^4$ K cm$^{-3}$. Can the starburst in
NGC~1705 keep this gas photoionized? Taking 
$n_e \sim$ 1 cm$^{-3}$, the measured
radius of the superbubble ($\sim$ 500 pc), and the
Lyman continuum luminosity of NGC~1705 ($Q$ = 10$^{52}$ s$^{-1}$),
the Stromgren thickness of the photoionized supershell will be
$\sim$10$^{21}$ cm and the column density of the ionized layer
is then $\sim$10$^{21}$ cm$^{-2}$. This is in good agreement
with the column density of photoionized gas we infer from the $FUSE$ data.
 
We can compare the basic dynamical properties of the warm ionized gas in
NGC~1705 to the predictions of the superbubble model. In convenient
units, the radius and expansion speed of the superbubble are given by:

\begin{equation}
r = 0.7 L_{mech,40}^{1/5} n_0^{-1/5} t_7^{3/5} kpc
\end{equation}

\begin{equation}
v = 41 L_{mech,40}^{1/5} n_0^{-1/5} t_7^{-2/5} km/s
\end{equation}

for an adiabatic superbubble inflated by a kinetic energy injection
rate $L_{mech,40}$ (in units of 10$^{40}$ erg s$^{-1}$) into
a uniform medium with nucleon density $n_0$ for a time $t_7$ (in units
of 10$^7$ years).

The age of the dominant super star cluster (NGC~1705-1) is well-constrained
to be about 10$^7$ years (de Mello, Leitherer, \& Heckman 2000).
This agrees with
the dynamical age ($t = 0.6 r/v$) of the superbubble
(Marlowe et al. 1995; Meurer et al. 1992).
The estimated kinetic energy injection rate due to supernovae is
$L_{mech,40}$ = 2 (Marlowe, Meurer, \& Heckman 1999). Finally, the HI column
in the center of NGC~1705 implies a mean nucleon density
of $n_0 \sim$1 cm$^{-3}$ for an ISM thickness of 1 kpc. 
Our estimate of the swept-up mass of gas in
the supershell (5$\times$10$^7$ $M_{\odot}$) divided by
volume enclosed by the supershell ($\sim5\times10^{64}$ cm$^{3}$)
likewise implies $n_0 \sim$1 cm$^{-3}$.
Equations 1) and 2) above then predict a radius of 0.8 kpc and an
expansion velocity of 50 km s$^{-1}$ for the supershell. Given the
overly-simplistic model (spherical symmetry, constant density),
we regard the agreement with the data as satisfactory.

\subsection{The Neutral Gas}

The neutral gas probed by $FUSE$ is significantly less blueshifted than the
ionized gas, and must therefore have a different origin.
For this gas to remain neutral, it must be optically-thick to
the Lyman continuum radiation from the starburst. This condition implies
that a slab of gas 
with a column density log$N$ = 20.2 located inside the superbubble
(and thus at a distance $r \leq$ 500 pc from an ionizing source with a Lyman
continuum luminosity
of $Q = 10^{52}$ s$^{-1}$) must have a density
$n >$ 8 cm$^{-3}$. 
Recall that the mean ISM density in NGC~1705 is roughly 1 cm$^{-3}$. Moreover,
log$N$ = 20.2 and $n >$ 8 cm$^{-3}$ implies that the
neutral absorbers must be very small (smaller than a few pc).

Thus, only relatively small dense clouds would be neutral. On-average,
the clouds are outflowing, so most can not be in any undisturbed ambient medium
in front of the superbubble's outer shell. We therefore
identify the outflowing
material with clouds in the superbubble's interior.
These were presumably clouds in the ISM of NGC~1705 that were 
overtaken and engulfed as the superbubble propagated   
through the more tenuous inter-cloud medium. This would be
a natural consequence of a multi-phase ISM (e.g., White \& Long 1991).

The dynamical model here is one of dense clouds exposed to the
outflowing hot gas in zones $2$ and $3$ of the superbubble. We would expect
to see
absorbing material that is injected from quiescent material at or near
$v_{sys}$, and which is then
accelerated up to some terminal velocity as it is transported outward
by the hot outflow.
Following Heckman et al. (2000), 
an interstellar cloud with column density $N$,
originally located a distance $r_0$ from a starburst, will be accelerated by
a spherically-symmetric outflow that carries an outward momentum flux $\dot{p}$
up to a terminal velocity given by:

\begin{equation}
v_{term} = 100 (\dot{p}/10^{32} dynes)^{1/2}(r_{0}/10^{21} cm)^{-1/2}
(N/10^{20} cm^{-2})^{-1/2} km/s
\end{equation}

We have chosen values for these parameters that are appropriate to the
neutral gas in NGC~1705. The predicted velocity range of absorption relative
to $v_{sys}$ would be $\sim$ 0 to -100 km s$^{-1}$. This agrees 
tolerably well with the velocities of the blueshifted
neutral gas observed in our $FUSE$ data.

\subsection{The Coronal Gas}

The detection of the interstellar
$OVI$ absorption-line (the first in a starburst galaxy)
is the most important result in this paper. We will organize our discussion
of the coronal gas in NGC~1705 as follows.

We begin by making some simple
inferences about the basic physical properties of the gas that are independent
of any specific model for the origin of the gas (\S~4.5.1).
In particular, in \S~4.5.2 we point out that - independent
of the detailed thermal/dynamical history of the gas -
the observed $OVI$ column density and line width imply that we are
observing gas that has been heated to $T \geq 3\times10^5K$ and which then
cools radiatively. These model-independent results can then be used
to constrain the radiative cooling rate from the coronal gas and show that
it is small compared to the supernova heating rate.   

We then turn our attention to specific models for the $OVI$. 
We will not exhaustively consider
all the possible alternative models (for example, conduction fronts associated
with either the supershell fragments or the HI clouds discussed above).
Instead we will describe the shortcomings of the
standard superbubble model (\S~4.5.3),
and then consider what we regard as an especially plausible alternative 
model in which the $OVI$ arises in a hydrodynamical
interaction between hot outrushing gas and the dense, cooler fragments 
of the superbubble shell during the ``blow-out'' phase. This idea is considered
analytically in \S4.5.4 in the context of models of turbulent mixing layers,
and then using numerical hydrodynamical models in \S4.5.5.

\subsubsection{Basic Physical Properties}

The combination of a detection of blueshifted $OVI\lambda$1032 in absorption
and an upper limit to corresponding redshifted emission allows us
to place an upper bound on the density and pressure in the coronal gas
(since we know the column density, and have an upper limit to the
emission-measure). This calculation only assumes reasonable symmetry between the
front and back sides of the outflow averaged over the kpc-scale
projected $FUSE$ aperture.

The $OVI$ ion is abundant only over a rather narrow temperature range
(e.g. Sutherland \& Dopita 1993), so
we will assume $T \sim$ 3$\times$10$^5$ K. 
The upper limit to the extinction-corrected $OVI\lambda$1032 flux
in the $FUSE$ aperture
corresponds to an upper limit to the $OVI$ intensity
of $I_{1032} \leq 2.5\times$10$^{4}$ photons cm$^{-2}$ s$^{-1}$ ster$^{-1}$.
Following Shull \& Slavin (1994), our observed $OVI$ column density
then leads to an upper limit to the electron density in the coronal
gas of $n_e \leq$ 0.1 cm$^{-3}$. The corresponding upper limit
to the pressure is $P/k \leq 6\times$10$^4$ K cm$^{-3}$. 

We have estimated a pressure in the warm photoionized gas of
$P/k \sim$ 2$\times$10$^4$ K cm$^{-3}$. The pressure in the coronal-phase gas
probed by the $OVI$ ion should be the same, since
zones $3$ through $5$ in the adiabatic superbubble model are isobaric
(MacLow \& McCray 1988). We note that the isobaric condition could be
invalidated by the presence of a
dynamically-significant magnetic field (in which case the thermal pressure
in the coronal gas should be higher than in the denser
photoionized material). However, our model-independent upper bound
of $P/k \leq 6\times$10$^4$ K cm$^{-3}$ in the coronal gas is only
a factor of three higher than the pressure implied by an 
assumption of isobaric conditions.
For $T \sim$ 3$\times$10$^5$ K,
our estimated pressure then implies that the characteristic density
in the coronal gas is $n_e\sim$3$\times$10$^{-2}$ cm$^{-3}$. For
a metal abundance of 1/8 Solar in NGC~1705 (Heckman et al. 1998)
the implied radiative cooling time from $T \sim$ 3$\times$10$^5$ K
is $t_{cool}$ = 1$\times$10$^6$ years (interpolating between the
collisional-ionization-equilibrium models for different
metallicities in Sutherland \& Dopita 1993). The assumption
of collisional-ionization-equilibrium is probably reasonable
for $OVI$, since the relevant recombination times at our assumed density
and temperature (Nahar 1999) are roughly an order-of-magnitude less than
the radiative
cooling time.

The estimated column density
in the coronal gas in NGC~1705 is $\sim$10$^{19}$ cm$^{-2}$,
so the above density implies a characteristic thickness for
the absorbing material of $\sim$ 100 pc. If this path length
is contributed by $N$ total clouds, the sound crossing time
of a cloud is $\sim1.2\times10^6 N^{-1}$ years. 
Thus, the ratio of sound-crossing
and radiative cooling times in a cloud is $\sim 1.2/N$ and the assumption
that the coronal gas is in rough pressure-balance with its surroundings
is therefore plausible.

We have estimated that the total mass of coronal-phase gas is
$\sim6\times10^5 M_{\odot}$. The above cooling time then
implies a cooling rate of $\dot{M} \sim$ 0.6 $M_{\odot}$ per year.
This can be compared to the average rate at which mass has been swept up
by the superbubble over its lifetime: $\dot{M} \sim M/t \sim 5\times10^7
M_{\odot}/10^7$ years $\sim$ 5 $M_{\odot}$ per year. 
The implied cooling luminosity is $\dot{E} = 3/2kT\dot{M}/\mu m_H \sim
10^{39}$ erg s$^{-1}$. This is about 5\% of the estimated
rate at which the starburst supplies mechanical energy.

Thus, radiative cooling associated with the coronal gas should not
dominate the dynamical evolution of the outflow. We note that radiative
cooling from the hotter gas detected in soft X-rays ($L_X \sim
10^{38}$ erg s$^{-1}$ - Hensler et al. 1998)
is significantly smaller than our estimate
of $\dot{E}$, and is therefore negligible.

\subsubsection{The Origin of the Observed Column Density}

Edgar \& Chevalier (1986) have computed the expected column
densities of various ions for the generic situation in which
gas is heated to a temperature $T_0 \sim 10^6$K and then cools
radiatively. The total column density of cooling gas 
is just given by $N_{cool} = \dot{N} t_{cool}$, where
$t_{cool}$ is the radiative cooling time and 
$\dot{N}$ is the rate of cooling per unit area. For a flow speed
$v$, mass conservation in the cooling flow implies
$v = \dot{N}/n_0$. Thus, the cooling column can also be 
written at $N_{cool} = n_0 t_{cool} v$.

It is important to note that $N_{cool}$ is independent of density,
since $t_{cool} \propto n_0^{-1}$. Moreover, at coronal temperatures
the cooling is dominated by metals, so $N_{cool} \propto t_{cool}
\propto Z^{-1}$ 
(where $Z$ is the metallicity). Since $N_{OVI} \propto N_{cool} Z$,
the characteristic $OVI$ column density in radiatively
cooling coronal gas is essentially independent of density and metallicity
and depends only on the value for $v = \dot{N}/n_0$.
Edgar \& Chevalier (1986) calculate that $N_{OVI,cool} \sim 4\times10^{14}$
cm$^{-2}$ ($v$/100 km/s). Calculations of $OVI$ production in high-speed
radiative shocks give similar values (Dopita \& Sutherland 1996).
This fiducial column density agrees well with our measured values
for $N_{OVI}$ and $v$ in NGC~1705.

This good agreement has two immediate implications. First,
independent of the detailed
dynamical and thermodynamical history of the $OVI$, this gas
it is almost certainly the result
of the radiative cooling of initially hotter gas.
Second,
we can now estimate the implied cooling rate.

Using the Edgar \& Chevalier (1986) models, 
our observed value $logN_{OVI}$ = 14.3 implies that
$\dot{N}/n_0$ = 5 $\times$ 10$^6$ cm s$^{-1}$. In their models the gas
cools from $T_0 =$ 10$^6$ K (and we note that significantly higher values
for $T_0$ 
in NGC~1705 are excluded by its low X-ray luminosity).
Since we estimate
$P/k$ = 2$\times$10$^4$ K cm$^{-3}$,
it follows that $n_0$ = 0.01 cm$^{-3}$ at $T_0 = 10^6$K and thus $\dot{N}$ =
5 $\times$ 10$^4$ cm$^{-2}$ s$^{-1}$. To calculate the implied rate at which gas
is cooling ($\dot{M} = \dot{N} A m_H$), we take a surface area
$A = 6 \times 10^{43}$ cm$^{2}$ for the NGC~1705 superbubble
(see above), and find that $\dot{M} = 0.07 M_{\odot}$ per year. As the gas
cools from 10$^6$ K its cooling luminosity is
$\dot{E} = 3/2kT\dot{M}/\mu m_H \sim
10^{39}$ erg s$^{-1}$. This is about 5\% of the estimated
rate at which the starburst supplies mechanical energy, which
agrees well with the more naive estimate
in \S~4.5.1 above. 

The above estimates of the cooling rate depend upon our assumed pressure.
However, the upper limit to the luminosity of $OVI\lambda$1032 emission-line
(\S~3.2.3) yields a pressure-independent upper limit to the cooling rates 
of $\dot{M} \leq$ 0.3 $M_{\odot}$ per year and
$\dot{E} \leq$ 4$\times$10$^{39}$ erg s$^{-1}$ (Edgar
\& Chevalier 1986). These limits are consistent with the above estimates.

\subsubsection{The Failure of the Simple Superbubble Model}

In the simple superbubble model there are two plausible origins
for the coronal gas. First, if the speed of outer shock driven into the
ambient ISM is high enough, $OVI$ ions will be abundant behind this shock
(in zone $5$).
This does not appear to be feasible in NGC~1705, since the minimum required
shock speed (post-shock temperature) is $\sim$ 150 km s$^{-1}$ 
(3$\times10^5$ K) - see Shull \& McKee (1979) and 
Dopita \& Sutherland (1996).
This is significantly higher than the expansion speed measured in
for the shell in NGC~1705: $\sim$ 53 km s$^{-1}$, corresponding
to a post-shock temperature of only 40,000~K.

Second, thermal conduction will transfer heat from zone $3$ to zone $5$
and create coronal-phase gas at the interface (zone $4$). Weaver et al.
(1977) predict an $OVI$ column density in this material of

\begin{equation}
N_{OVI} = 1\times 10^{14} Z_O n_0^{9/35} L_{mech,40}^{1/35} t_7^{8/35} cm^{-2}
\end{equation}

where $Z_O$ is the oxygen abundance relative to Solar. For $Z_O$ = 1/8,
the predicted value for $N_{OVI}$ is a factor of $\sim$15 smaller than
we measure. In this case, $N_{OVI}$ is smaller than the 
value for radiatively cooled gas (\S~4.5.2) and has a
direct dependence on the metallicity. This is because cooling
in the conductively heated zone is dominated by adiabatic expansion losses
(Weaver et al. 1977).

The relative velocity of $OVI$ absorption in NGC~1705
is also inconsistent with the simple superbubble model, which
predicts that the outflow velocity in this material (zone $4$) will 
be substantially less than 
the outflow speed of the superbubble shell (zone $5$). Our data instead
show that the $OVI$ outflow speed is probably even larger than that
of the shell (77$\pm$10 vs. 53$\pm$10 km s$^{-1}$ - see above).

We conclude that the simple superbubble model does not account
for the observed properties of the $OVI$ absorption-line in NGC~1705.

\subsubsection{Hydrodynamical Heating During Blow-Out}

We have argued in \S~4.5.2 that the $OVI$ arises in a flow of gas that
has been heated to an initial temperature of $T_0 \geq 3\times10^5K$
and which is then cooling radiatively. As discussed in \S~4.5.3,
this heating/cooling process does not correspond to the outer shock
in the superbubble (zone 5) because the observed expansion velocity
of the superbubble is too slow to produce $OVI$ behind this shock.
In this section, we therefore consider a plausible alternative model.

We have argued above that that the superbubble in NGC~1705
is in the process of breaking out of the ISM of NGC~1705. During this phase,
the expansion speed of the superbubble shell accelerates and Rayleigh-Taylor
instabilities will cause the shell (zone $5$) to fragment. This
allows the hot X-ray emitting gas in zone $4$ to push its way out
through the fragmented
shell, and Kelvin-Helmholtz instabilities are believed to lead
to the turbulent mixing of this hot outrushing gas with the
cooler shell fragments (e.g. MacLow, McCray, \& Norman 1989).
Slavin, Shull, \& Begelman (1993 hereafter SSB) have investigated
the emission and absorption-lines produced by the intermediate-temperature
(coronal phase) gas in these ``turbulent mixing layers'' (TMLs).

We emphasize that the
kinematics of the $OVI$ absorbers in NGC~1705 are consistent with
TMLs. The blueshift of the $OVI$ relative to the shell material
is expected as the hot gas rushes out through ``cracks'' in the shell.
The magnitude of this blueshift follows from SSB:
the velocity of the intermediate temperature gas
in the TML relative to the cool gas from which it is created
is given by $v_{tml} \sim v_{hot}(T_{cool}/T_{hot})^{1/2}$, where
$v_{hot}$ is the relative flow speed between the gas at $T_{cool}$
and the outrushing hot gas at $T_{hot}$. For the photoionized shell
fragments that dominate the mass in NGC~1705 $T_{cool} \sim 10^4$ K.
Based on X-ray spectroscopy of dwarf starburst galaxies (della Ceca et
al. 1996,1997; Hensler et al. 1998),
$T_{hot} =$ few$\times$10$^6$ to 10$^7$ K. As this hot gas
flows past the shell fragments, its maximum relative velocity will be
roughly its sound speed (i.e. $v_{hot} \sim$ 500 km s$^{-1}$).
Thus,
for $T_{cool}/T_{hot} \sim$ 10$^{-3}$, $v_{tml} =$16 km s$^{-1}$ for
$v_{hot}$ = 500 km s$^{-1}$. This can be compared with
NGC~1705 where the $OVI$ absorption line is blueshifted by
24$\pm$10 km s$^{-1}$ relative to the photoionized shell material. 

As pointed out by SSB, the ionic column density per 
TML is independent of the pressure in the gas, but is directly
proportional to the relative velocity of the cool and hot gas
($v_{hot}$). Also the $OVI$ column will be appreciable only if the maximum
temperature of the TML exceeds $\sim3\times10^5$ K.
Their models only extend up to $v_{hot}$ = 100 km s$^{-1}$ and $logT_{tml}$
= 5.5,
and these predict $OVI$ columns that are of-order 10$^{12}$ cm$^{-2}$
per TML. These values could be adjusted upward 
in TMLs with larger values for $v_{hot}$ and $logT_{tml}$.
In terms of the arguments in \S~4.5.2, the low $OVI$ columns
in the TML models are partly attributable to the small relevant
flow velocity in the TML:
$\dot{N}/n_0 = v_{tml} = v_{hot}(T_{cool}/T_{hot})^{1/2} \sim$ 10 km s$^{-1}$. 
It appears that of-order 10$^2$ TMLs per line-of-sight are needed
in NGC~1705 (we see a `sea'' of shell fragments, each with its own TML).
Proportionately fewer TMLs would be needed for larger flow velocities.

\subsection{Insights from Numerical Hydrodynamics}

As an initial attempt to investigate qualitatively 
the dynamics of gas in the complex
transitional stage between superbubble and superwind, we have performed several
2-dimensional hydrodynamical simulations of a superbubble blowing
out of a dwarf galaxy. We calculate the column densities and velocity
structure of simulated lines of sight through these models,
in particular the relative velocities of
the warm and coronal phases that the $FUSE$ observations of NGC~1705 show
can not be explained by the classic superbubble model.

The simulations were performed using the hydrodynamical code described
in Strickland \& Stevens (2000), with the initial conditions altered to
roughly correspond to NGC~1705. These simulations were performed in
cylindrical coordinates, assuming rotational symmetry around the $z$-axis
(the minor axis of the galaxy). The hydrodynamical grid covers a
region 2.5 kpc in radius by 2.5 kpc high along the $z$-axis with
500 by 500 equally sized cells.

The initial ISM was set up in rotating hydrostatic equilibrium.
We use a King model for the gravitational
potential, producing a rotational velocity
at $r = 7$ kpc of $65$ km s$^{-1}$. The ambient gas was given a temperature of
$1.6 \times 10^{4}$ K, to approximate turbulent pressure support
of the ISM, based on the $\sim 15$ km s$^{-1}$ velocity dispersion of
the HI gas. The resulting initial ISM density distribution has a
peak number density (in the nucleus) of $2.5$ cm$^{-3}$,
a total column density through the galaxy along the minor axis of
$N_{\rm H} = 3.5 \times 10^{21}$ cm$^{-2}$, 
a vertical scale height of 0.2 kpc, and a total mass
of $M_{\rm gas} = 8 \times 10^{7} M_{\odot}$ (within $r = 2.5$ kpc).
These properties are all crudely similar to those observed in
NGC~1705 (Meurer et al. 1998).

We modeled the return of mass and mechanical energy into the ISM from 
starburst event as a single instantaneous burst
of star-formation, using the low metallicity models ($Z = 0.25 Z_{\odot}$)
of Leitherer \& Heckman (1995). Appropriate amounts of mass and
thermal energy were added at each computational time step to
those cells within our assumed starburst region, a cylindrical region
$150$ pc in radius and $60$ pc thick vertically.

Our fiducial model has a time-average mechanical luminosity of
$L_{mech} = 2 \times 10^{39}$ erg s$^{-1}$, somewhat weaker than
the observed starburst NGC~1705, but chosen to give blow-out
at $t \sim 13$ Myr (approximately the observed dynamical age of
the superbubble in NGC~1705). At blow-out the radius of superbubble is
$\sim 500$ pc in the plane of the galaxy, and about twice as large
along the minor axis.

We assume an ISM metal abundance of
1/10 Solar for the purposes of calculating radiative cooling
and column densities. We also performed other simulations to crudely
explore parameter space: First, a simulation with a more powerful
starburst ($L_{mech} = 7 \times 10^{39}$ erg s$^{-1}$), which
blows out earlier at $t \sim 8$ Myr but otherwise is very similar
to the fiducial model in other properties. 
Second, a model using Solar abundance to investigate the scaling of
column density with metal abundance. Third, a higher resolution
model with twice the resolution in every dimension, to investigate
the dependence of calculated properties on numerical resolution.

An example of the model is shown in Figure 4, which plots the density
and temperature of the gas in our fiducial low-metallicity model
at t = 12.5 Myr.
The $OVI$ absorption seem in these simulations arises
primarily in material at the interfaces between cool dense gas, e.g. in the
shell of the superbubble or shell fragments after blow-out,
and the hot ($T>10^6$ K) gas filling the bubble volume
(see Figure 4). In almost all cases these interfaces are numerically
unresolved. 

\begin{figure}
\includegraphics[height=7cm]{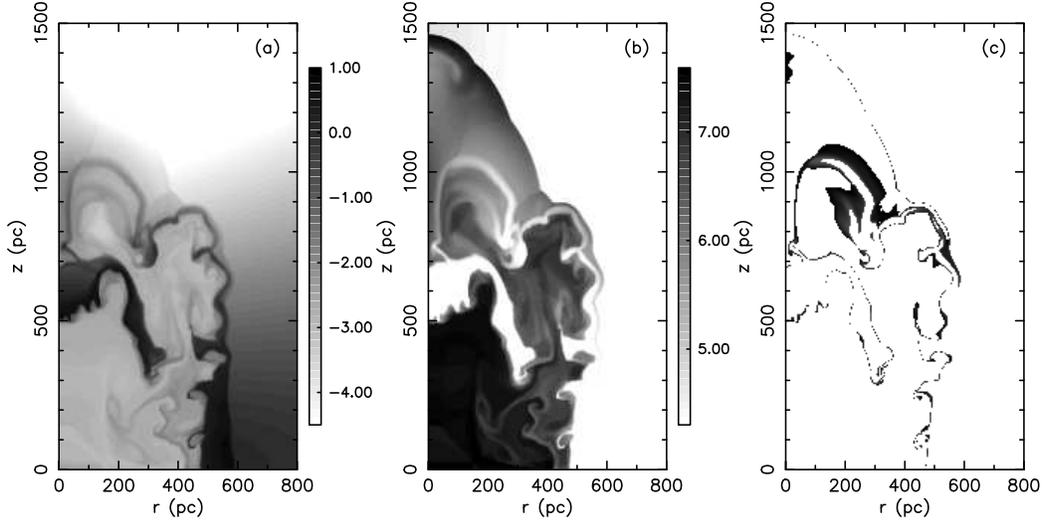}
\caption{Grey scale images of gas density and temperature in our
fiducial hydrodynamical simulation of a dwarf starburst, shown
as the superbubble blows out of the disk at $t=12.5$ Myr.
The three panels show (a) log number density
(cm$^{-3}$), (b) log gas temperature (K) and (c) the location of
the $OVI$-absorbing material (gas with $5.2 \le$ log T (K) $\le 5.8$).}
\end{figure}

\begin{figure}
\includegraphics[height=7cm]{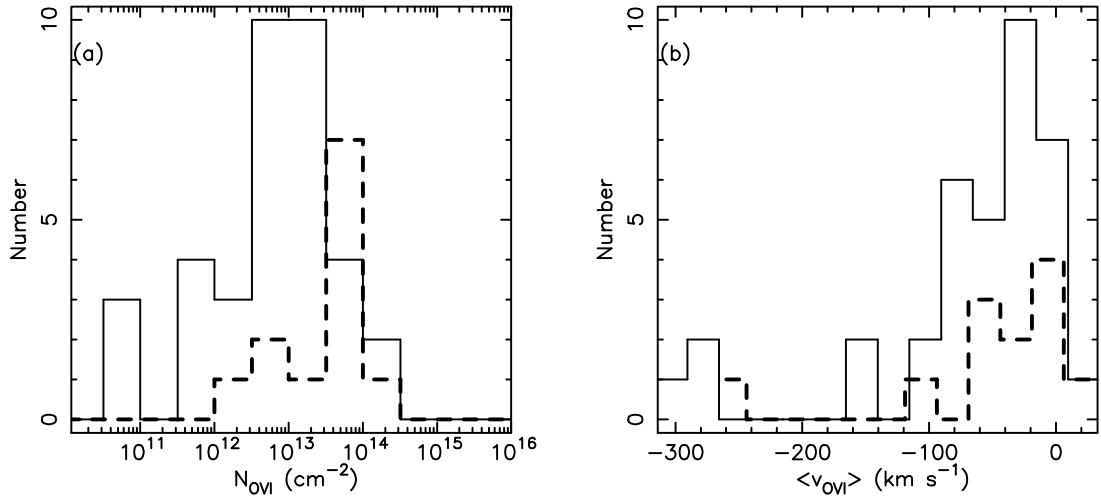}
\caption{(a) Histograms of $OVI$ column density, covering
all lines of sight and all epochs in the three low metallicity
numerical simulations (solid  line) and the simulation using Solar
metallicity (dashed line).
(b) Corresponding histograms of the mean $OVI$ absorption line velocity
(weighted by the $OVI$
column density).}
\end{figure}

We calculate the column density as a function of
line-of-sight velocity profiles in 
each model at three different epochs (pre-, during, and post- blow-out,
typically at $t = 10, 13$ and $18$ Myr respectively),
for four different lines of sight from the nuclear starburst
(assuming galaxy inclination angles of ~$i = 0, 30, 60$ \& $80\degr$),
and for two different ionic species that probe
different observed phases in superwinds (warm gas: $NII$; coronal gas: $OVI$).
 Fractional ion abundances as a function of temperature
were taken from Sutherland \& Dopita (1993), which assume collisional ionization
equilibrium. This a reasonable assumption in the case of $OVI$ (see above).
It is also a reasonable
mock-up of $NII$ (in reality, photoionization by stars in NGC~1705 keeps the
gas containing $NII$ near $T \sim$ 10$^4$, close to the 
temperature floor of 1.6 $\times$10$^4$ K in the simulations).

Much of the UV light in NGC~1705 comes from the compact central star cluster, 
but a significant fraction comes from a more diffusely spread 
component (Meurer et al 1995). The width of $30\arcsec$ FUSE aperture used
corresponds to a projected width of $\sim 900$ pc at the distance of NGC 1705,
so the observed spectrum is a sum over many slightly different lines of
sight. To investigate the effect of this on the simulated absorption line
profiles we explored two cases, which should bracket the real situation.
In the first case we assumed all the light comes from a single point-like
source at the center of the galaxy, and in the second case that the light
comes from uniform surface brightness disk of radius $300$ pc.
We found very little difference in the simulated absorption line
profiles between the two methods, and so quote only results 
using the point-like light source.

For any particular model, the column densities and kinematics
of the gas vary strongly between different sight lines, due to
the structural complexity that arises in 2-D simulations (see Suchkov
et al. 1994, Strickland \& Stevens 2000).
The total column densities (in either $OVI$ or $NII$) do not
change systematically with time in these simulations.
We find that the mean and variance 
in total column density is the same before, during and after blow-out,
considering all 12 simulated lines of 
sight at each specific epoch.

As shown in Figure 5, we find $OVI$ column
densities in the range $\log N_{OVI} \sim 13\pm{0.5}$ in the low metallicity
models ($Z = 0.1 Z_{\odot}$), and $\log N_{OVI} \sim 14\pm{0.5}$
in the Solar metallicity model. Only about 10\% of the lines-of-sight 
in the low metallicity models have $OVI$ columns comparable to the
observed value in NGC~1705.
One reason for the failure of these simulations to reproduce the
observed OVI column densities may be that there is insufficient
numerical resolution to resolve the cooling zones between hot and cool
gas, where the OVI-absorbing gas exists. Evidence to support this idea
comes from comparing the predicted column densities in the 
fiducial model with those in the high resolution model. 
Although the predicted columns for low ionisation species always agree 
to within a factor 2, the predicted OVI column densities agree to 
within a factor 2 in only 50\% of the sight lines, with the remaining
sight lines differing by a factor of up to 30. 
The lack of thermal conduction in these models may also be partially
responsible for the low OVI columns, although the arguments presented 
in \S~4.5.3 suggest thermal conduction alone can not produce
the observed OVI columns. Our simulations do show that 
the column density of lower ionization
material is considerably higher than that of coronal material,
and that this phase contains most of the mass in the 
outflow (consistent with our observations).

The column-weighted mean velocities for the coronal gas along the
different sight-lines
generally lie in the range $v_{OVI} \sim -100$ to $\sim 0$ km s$^{-1}$,
although the full range covers $v_{OVI} \sim -320$ to $\sim +50$ km s$^{-1}$
(Figure 5).
This gas is typically more blue shifted, and has a wider range of outflow
velocities, than the cooler gas in our simulations.
For example, the typical velocity range for $NII$ is 
$v_{NII} \sim -70$ to $\sim 0$ km s$^{-1}$ and the 
full range seen only covers $v_{NII} \sim -110$ to $\sim0$ km s$^{-1}$.
The velocity range covered appears very similar in both low and Solar
metallicity models, so it appears that the kinematics of the gas in these
simulations is independent of the metal abundance.
For $OVI$, the mean outflow velocity (and range of velocities), 
does appear to increase with time, moving from pre-blow-out through
to post-blow-out. This is not the case for lower ionization 
species, e.g. the distribution of mean velocity for $NII$ is very similar
at all three epochs.

%Column density against velocity profiles for individual sight lines
%for epochs prior to and during k out typically show a single
%dominant component, for both cool and coronal phases.
%This component covers only a narrow range in velocity ($\sim 25$ to
%$50$ km s$^{-1}$). This suggests we see only a single object
%along the line of sight at these epochs. Complex, multi-velocity
%component profiles only appear after break out in these
%models, and only in higher ionization
%species (OVI). Note that this may be an effect of numerical resolution,
%as the coolest (hence densest and most compact) 
%material is the least well resolved numerically, and hence the number of
%small dense clouds produced in the simulation is strongly
%resolution dependent.

In summary, while the numerical models underpredict the $OVI$ columns, they
{\it can} explain one aspect of the observed
data on NGC~1705. The mean outflow velocity of high-ionization
gas probed by $OVI$ is as high as, or higher than, that of low-ionization gas.
This is in contrast with the standard superbubble model
in which the outflow speed in the coronal gas (zone $4$) is less than that
of the superbubble shell. This result supports
the idea that blow-out is responsible for the kinematics
of the coronal phase in NGC~1705.

\section{Implications}

As summarized in \S~1, probably the greatest uncertainty in assessing
the impact of starburst-driven outflows on the evolution of dwarf
galaxies is the importance of radiative cooling (which will directly
determine the outflow's dynamical evolution). In the absence of severe radiative
cooling, both analytic arguments and hydrodynamical simulations imply
that outflows like the one in NGC~1705 will easily blow-out
of their galaxy's ISM
(e.g., MacLow \& McCray 1988; Marlowe et al. 1995;
Martin 1998,1999;
MacLow \& Ferrara 1999; Strickland \& Stevens 2000). The hot gas
that blows out of the superbubble will then
be able to heat and chemically enrich the inter-galactic
medium (e.g., Ponman et al. 1999; Tozzi, Scharf, \& Norman 2000)
and/or an extended gaseous halo
surrounding the dwarf galaxy (Silich \& Tenorio-Tagle 1998).

We will not repeat
these arguments here, but simply note that within the context
of the standard McKee \& Ostriker (1977) formalism, the lack of
radiative cooling would be a result of the exceedingly high supernova rate
per unit volume, which leads to a porosity near unity in the ISM
(i.e., the ISM within which the supernovae detonate is filled
almost entirely with hot, diffuse gas). This is especially pertinent
to NGC~1705, in which nearly half of the supernovae will detonate
inside the pc-scale super-star-cluster NGC~1705-1.

Existing X-ray data establish that radiative cooling from hot
($T \geq 10^6$K) gas is energetically insignificant in NGC~1705
(Hensler et al. 1998) and other starburst galaxies (cf. Heckman 2000
and references therein).
The new data from $FUSE$ are very instructive, 
since they directly probe for the coronal-phase 
($T \sim 10^5-10^6$ K) gas near the peak of the cooling curve
where hitherto-unobserved radiative cooling might be severe.
As we showed in \S~4.5.2 above,
the observed $OVI$
column density ($1.8\times10^{14}$ cm$^{-2}$) and flow speed
($\sim10^2$ km s$^{-1}$) implies that we are seeing gas
in the outflow that has
been heated to $T \geq 3\times10^5$ K and then radiatively cooled.
We have estimated that the implied rate of radiative
cooling is only $\sim$6\% of the rate of supernova heating
(\S~4.5.2). Thus, there is no place left to hide the radiation necessary to
quench the supernova-driven outflow in NGC~1705.

What then is the fate of the outflowing gas? 
For an isothermal gravitational potential that extends to a maximum radius
$r_{max}$, and has a virial velocity $v_{rot}$, the escape velocity at a
radius $r$ is given by:

\begin{equation}
v_{esc} = v_{rot}[2(1 + ln(r_{max}/r))]^{1/2}
\end{equation}

The rotation curve for NGC~1705 implies $v_{rot}$ = 62 km s$^{-1}$
(Meurer et al. 1998), and the rotation curve remains relatively
flat out to the last measured point at $r$ = 7 kpc. Thus, the minimum
escape velocity at the present location
of the superbubble wall ($r \sim$  0.5 kpc) will be roughly 170 km s$^{-1}$.
This is significantly higher than the velocities of most of the outflowing gas
that we observe in-absorption in
NGC~1705. Thus, there is no direct evidence that most of this material
will escape NGC~1705. Maps of the HI in NGC~1705 (Meurer et al. 1998)
show that the large-scale ISM in NGC~1705 will not be strongly affected
by the outflow (which will only puncture a few-kpc-scale hole in the
central region of the HI disk).

On the other hand, the bulk of kinetic
energy and most of the newly-created metals will reside in the hotter
gas that is driving the superbubble's expansion. The temperature of this
gas will be $>10^6$ K (della Ceca et
al. 1996,1997; Hensler et al. 1998) . We have argued that
we are witnessing the blow-out phase in NGC~1705 in which this gas
is vented through the fragmented superbubble wall. Following blow-out,
in the absence of significant radiative cooling
this hot gas will feed a galactic wind with a terminal velocity
of at least 300 km s$^{-1}$ (Chevalier \& Clegg 1985). This
is comfortably above $v_{esc}$ for any plausible value for $r_{max}$.
Once the hot gas vents out of the ruptured superbubble, there is
no known external medium to impede the resulting high-speed wind.

We conclude that NGC~1705 can eject most of the newly-created metals
and kinetic energy returned by the central starburst over the past
$\sim$ 10$^7$ years. If typical of dwarf starbursts, this process
could help create the low metal abundances in dwarf galaxies
(e.g., Dekel \& Silk 1986; Lyden-Bell 1992) and contribute to the
heating and chemical enrichment of the intergalactic medium
(Ponman et al. 1999; Gibson et al. 1997; Ellison et al. 2000).

\section{Conclusions}

We have presented new $FUSE$ far-UV spectroscopy of the prototypical
dwarf starburst galaxy NGC~1705. Previous optical and UV spectroscopy
has established that the starburst in this galaxy is driving a
large-scale outflow (Meurer et al. 1992; Marlowe et al. 1995).
The $FUSE$ data are especially important because
they probe the coronal-phase ($T \sim$ 10$^5$ to 10$^6$ K) gas
that may dominate the radiative cooling of the outflow, and thereby
largely determine its dynamical evolution. The high quality of the data
(spectral resolution of $\sim$ 30 km s$^{-1}$ and a $S/N \sim$ 16)
gives us important new insight into all the phases in the ISM of this
galaxy.

Firstly, we do not detect any $H_2$, with an upper limit
of $logN_{H2} \leq$ 14.6. This is similar to the results obtained
by Vidal-Madjar et al. (2000) for the dwarf starburst IZw18.
In the specific case of NGC~1705 the low $H_2$ abundance
is consistent with Galactic sight-lines with similarly low
$E(B-V)$ (Savage et al. 1977).

We do detect absorption from three other phases of the ISM:
neutral gas, warm photoionized gas, and coronal-phase gas
(probed with the $OVI\lambda$1032 line).
The total column densities (cm$^{-2}$) in the three phases are $logN$
= 20.2, 20.9, and $\geq$19.0 respectively. The first is directly measured
from Lyman series, while the latter two are estimated from
the measured ionic columns assuming a metal abundance of 12\% solar
(as measured in the optical emission-line gas).

All the interstellar absorption-lines associated with NGC~1705
are broad ($\sim$ 10$^2$ km s$^{-1}$) and blueshifted. Thus,
gas is flowing out of NGC~1705 with
$(v-v_{sys})$ = -32, -53, and -77 km s$^{-1}$ for the neutral,
photoionized, and coronal gas respectively.

The mass and kinetic energy in the outflowing  gas probed by $FUSE$ is dominated
by the warm photoionized gas, which is also seen
via its optical line-emission. The size, morphology, outflow speed,
mass, kinetic energy, and dynamical age of this structure are all consistent
with a
simple model of an adiabatic superbubble whose expansion
is driven by a piston of hot gas created by the cumulative effect
of the supernovae in the starburst (e.g., Weaver et al. 1977).

In contrast, we have shown that the neutral gas must reside in relatively
small dense
clumps in order to be optically-thick in the Lyman continuum.
We attribute the outflowing material to clouds in the ISM that have been
overtaken by the superbubble and accelerated by 
the outflowing hot gas. 

The properties of the coronal gas probed with $OVI\lambda$1032 are
inconsistent with the predictions of the simple superbubble model.
The observed expansion speed of the superbubble (53$\pm$10 km s$^{-1}$)
is too small to produce $OVI$ behind its shock-front. The blueshift
of the $OVI$ line is larger than that of superbubble shell (while
the model predict that it should be substantially smaller).
Finally, the observed $OVI$ column ($logN_{OVI}$ = 14.3) is a factor of 
$\sim$15 too large compared to the superbubble model (in which such gas
arises in a conductively-heated and adiabatically-cooled interface
between hot gas and the cool outer superbubble shell).

We therefore proposed the following origin for the $OVI$ absorber,
and have investigated its plausibility via numerical hydrodynamical models.
The morphology and relative size-scale of the superbubble implies that
it has begun to ``break out'' or ``blow out'' of the ISM in the disk
of this galaxy. During the blow-out phase, the superbubble shell
will accelerate and then fragment (e.g., MacLow \& McCray
1988). As it fragments, the piston
of hot gas inside the superbubble will flow out through the
``cracks'', and the resulting hydrodynamical interaction
between this outrushing gas and the shell fragments will
create intermediate-temperature
coronal gas that can produce the observed $OVI$ absorption
(e.g., Slavin et al. 1993). This process accounts nicely
for the observed kinematics of the $OVI$ absorption. We also
emphasized that for the observed flow speed of $\sim10^2$ km s$^{-1}$,
the observed $OVI$ column density is just what is expected for
gas that has been collisionally-ionized and then radiatively-cooled,
independent of its density or metallicity and independent
of any specific hydrodynamical model (Edgar \& Chevalier 1986).

We have argued that the coronal gas should be in rough pressure-balance
with the warm photoionized gas (whose pressure we determined to
be $P/k=2\times10^4$ K cm$^{-3}$). This allowed us to estimate
cooling rates of $\sim$0.07 $M_{\odot}$ per year and
$\sim$10$^{39}$ erg s$^{-1}$ in the
coronal gas, based on the $OVI$ absorption-line. Independent
of the gas pressure, the lack of redshifted $OVI$ {\it emission}
from the backside of the outflow implies upper limits to the
cooling rates of $\leq $0.3 $M_{\odot}$ per year and
$\leq 4\times10^{39}$ erg s$^{-1}$. The rate of radiative cooling due to
hotter ($T \geq 10^6$ K) gas is negligible in comparison
(Hensler et al. 1998).
Since the rate of supernova-heating in NGC~1705 is
$2\times10^{40}$ erg s$^{-1}$ (Marlowe et al. 1999), we have argued
that radiative cooling is not dynamically important
in the NGC~1705 outflow.

In the absence of significant radiative cooling, the superbubble
in NGC~1705 should be able to blow out of the ISM. Based
on the relatively low outflow velocities, it is unlikely
that the absorption-line gas we observe with $FUSE$ will escape
from the galaxy. Likewise, the bulk of the global ISM ($HI$)
in NGC~1705 will be retained (cf. De Young \& Heckman 1994; MacLow
\& Ferrara 1999).
However, we have argued that
the hotter gas that drives the outflow (and which is now being vented
out of the ruptured superbubble) can escape and thereby
carry away most of the
kinetic energy and newly-created metals supplied by the starburst.
This process
has potentially important implications for the evolution of both dwarf
galaxies and the IGM.

\acknowledgments

We thank the members of the $FUSE$ team for providing this superb
facility to the astronomical community. We thank Colin Norman and Robin
Shelton for enlightening discussions, and an anonymous referee
for a detailed and thoughtful report that improved the paper.
This work was supported in part
by NASA grants NAG5-6400 and NAG5-9012. DKS
is supported by NASA through {\it Chandra} Postdoctoral Fellowship Award
Number PF0-10012, issued by the {\it Chandra} X-ray Observatory Center,
which is operated by the Smithsonian Astrophysical Observatory for and on
behalf of NASA under contract NAS8-39073.

\clearpage

\begin{deluxetable}{lrrrrr}
\tablecolumns{6}
\tablenum{1}
\tabletypesize{\scriptsize}
\tablecaption{Equivalent Widths and Radial Velocities}
\tablewidth{0pt}
\tablehead{
\colhead{ID} & \colhead{$\lambda$}   & \colhead{$W_{MW}$}   &
\colhead{$W_{N1705}$} &
\colhead{$v_{MW}$}  & \colhead{$v_{N1705}$} \\
\colhead{(1)}&\colhead{(2)}&
\colhead{(3)}&\colhead{(4)}&
\colhead{(5)}&\colhead{(6)}
}
\startdata
H2 Q(1)& 1009.770  &     $<$35(3$\sigma$) & $<$35(3$\sigma$) &  ...  &  ... \\

C II   &1036.337   &    510$\pm$30 (L1) & $>$680 (L1)c     &  +18   &   +582 \\
       &           &    492$\pm$22 (L2) & $>$670 (L2)c     &       &        \\

C III  & 977.020   &    560$\pm$30 (S2) & $>$800 (S2)b     &  +28   &   +555 \\
       &           &    520$\pm$35 (S1) & $>$700 (S1)b     &       &       \\

N I    &1134.165    &   121$\pm$10 (L1)  & 23$\pm$10 (L1)   & +28     & +582 \\
       &            &   134$\pm$10 (L2) &  24$\pm$10 (L2) &        &      \\

N I    &1134.415    &   155$\pm$10 (L1)  & 32$\pm$10 (L1) &   +24   &   +590 \\
       &            &   156$\pm$10 (L2)  & 23$\pm$10 (L2) &        &        \\

N I    &1134.980    &   171$\pm$11 (L1)  & 35$\pm$11 (L1) &   +25   &   +602 \\
       &            &   161$\pm$10 (L2)  & 35$\pm$10 (L2) &        &        \\

N II   &1083.990   &    356$\pm$25 (S2) & 338$\pm$35 (S2)  &  +7  &    +573 \\

O I    &1039.230    &   260$\pm$20 (L1)a &  243$\pm$15 (L1) &  +14 &  +592 \\
       &            &   252$\pm$20 (L2)a &237$\pm$15 (L2)   &      &       \\

O I    &988.773     &  302$\pm$25 (S2)a & 480$\pm$30 (S2) &  +10   &   +595 \\
       &            &   283$\pm$30 (S1) & 540$\pm$45 (S1) &        &        \\

O VI   &1031.926    &   252$\pm$22 (L1) & 171$\pm$18 (L1) &   +25   &   +545 \\
       &            &   220$\pm$22 (L2) & 191$\pm$19 (L2)   &        &        \\

Si II  &1020.699    &   129$\pm$16 (L1) & 94$\pm$24 (L1)   &  +16   &  +589 \\
       &            &   104$\pm$12 (L2) &  65$\pm$20 (L2)   &       &       \\

S III  & 1012.502    &   143$\pm$20 (L1)b &  250$\pm$15 (L1) &  +8 &   +578 \\
       &            &   128$\pm$20 (L2)b & 215$\pm$20 (L2)   &      &       \\

S IV   &1062.662   &    $<$36 (3$\sigma$) &    85$\pm$15 (L1) & ... &  +555 \\
       &           &     $<$36 (3$\sigma$)&    82$\pm$15 (L2) &     &       \\

Ar I   & 1048.220  &     76$\pm$10 (L1) &  60$\pm$10 (L1)   &  +20   &   +566 \\
       &            &    83$\pm$10 (L2) &  80$\pm$20 (L2)   &       &       \\

Fe II  &1063.176    &   125$\pm$11 (L1) & 162$\pm$16 (L1)   & +11    &  +597 \\
       &            &   140$\pm$13 (L2) & 177$\pm$21 (L2)   &       &       \\

Fe II  &1144.938    &   192$\pm$12 (L1) & 262$\pm$17 (L1)    & +20   &   +601 \\
       &            &    206$\pm$11 (L2)&  241$\pm$17 (L2)  &     &        \\

\enddata

%% Text for table notes should follow after the \enddata but before
%% the \end{deluxetable}. Make sure there is at least one \tablenotemark
%% in the table for each \tablenotetext.
\tablecomments{The line equivalent widths in the Milky Way (Col. 3)
and NGC~1705 (Col. 4) are in m\AA. These widths and the 1$\sigma$ errors
were derived according to the 
prescription outlined by Sembach \& Savage (1992).  The values account for 
statistical noise and modest continuum placement uncertainties.  In some cases,
additional uncertainties (not listed) due to stellar blending and 
fixed-pattern noise introduced by the FUSE detectors may be warranted.
The detector segment utilized
(LiF1, LiF2, SiC1, and SiC2) is indicated in parentheses
following each equivalent width. 
The letters appended after these measurements have the following meaning:
$a$ - Value may be affected by terrestrial O I airglow emission;
$b$ - Line probably has some stellar blending, measurement uncertain;
$c$ - Lower limit because of blending with other lines.
The radial velocities of the centroids for Milky Way (Col. 5)
and NGC~1705 lines (Col. 6) are in the LSR frame. These were derived by
fitting single Gaussian profiles to the absorption lines.
Typical uncertainties are
$\pm$10 km s$^{-1}$. The wavelength scale in our data has been
adjusted so that the Milky Way neutrals agree with the mean velocity of the 
lines from similar species in our $HST$ STIS echelle spectrum of NGC~1705
(Sembach et al. 2001; Heckman et al. 2001).
}
\end{deluxetable}

\clearpage

\begin{deluxetable}{lrrrrr}
\tablecolumns{6}
\tablenum{2}
%\tabletypesize{\scriptsize}
\tablecaption{NGC~1705 Column Densities}
\tablewidth{0pt}
\tablehead{
\colhead{Species} & \colhead{$\lambda$}   & \colhead{$log(f\lambda)$} &
\colhead{$logN$} &
\colhead{$logN$} &
\colhead{$logN$}\\
\colhead{}&\colhead{}&
\colhead{}&\colhead{(Side1)}&
\colhead{(Side2)}&
\colhead{Final}
}
\startdata

H2 (J=0) &              &          & $<$14.2      &     $<$14.2 & $<$14.2 \\
H2 (J=1) &              &          & $<$14.2      &     $<$14.2 & $<$14.2 \\
H2 (J=2) &              &          & $<$14.2      &     $<$14.2 & $<$14.2 \\

H I      &      1025.7  &  1.909   &    20.2     &     20.2 & 20.2$\pm$0.2\\ 

N I      &      1134.1  &  1.182   &    $<$14.17    &   $<$14.30 & $<$14.30 \\
         &      1134.4  &  1.483   &    $<$14.08    &     $<$14.07 & $<$14.07 \\
         &      1135.0  &  1.660   &    13.97     &     13.98 & 13.97$\pm$0.08\\

N II     &      1084.0  &  2.048   &    ---       &    14.70 & 14.70$\pm$0.15 \\

O I      &      1039.2  &  0.980   &    15.64     &    15.63 & 15.63$\pm$0.08 \\

O VI     &      1031.9  &  2.137   &    14.26     &    14.26  & 14.26$\pm$0.08\\

Si II    &      1020.7  &  1.460   &    14.68     &    14.51 & 14.61$\pm$0.14 \\

S III    &      1012.4  &  1.556   &    15.02     &    15.00 & 15.02$\pm$0.12 \\

S IV     &      1062.7  &  1.628   &    14.45     &    14.43 & 14.45$\pm$0.08 \\

Ar I     &      1048.2  &  2.408   &    13.44     &    13.49 & 13.46$\pm$0.10 \\

Fe II    &      1144.9  &  2.084   &    14.53     &   14.54 &  14.54$\pm$0.10 \\
         &      1063.2  &  1.805   &    14.54     &   14.55 &  14.54$\pm$0.10 \\

\enddata
\tablecomments{The $f$-values are taken from Morton (1991), and $\lambda$
is in \AA. 
Column densities (cm$^{-2}$) were derived 
by converting the observed absorption
profiles into optical depth profiles and integrating over velocity (see
Savage \& Sembach 1991). These 
column densities do not have any saturation corrections applied, though these 
are expected to be small given the great breadth of the lines.  A comparison 
of the results for the two $FeII$ lines listed indicates that this is 
indeed the case.
In the case of the $HI$ column, we quote the average of the value
obtained from the higher-order Lyman series in the $FUSE$ data
and the value derived from $Ly\alpha$ by Heckman \& Leitherer (1997).
}
\end{deluxetable}

\end{document}